\documentclass[showpacs,preprintnumbers,amsmath,amssymb,twocolumn]{revtex4}
 \usepackage{color}
\usepackage{graphicx}% Include figure files
\def\<{\langle}
\def\>{\rangle}
\def\tr{{\rm{tr}}}

\def\i{\textmd{i}}

\def\R{{\mathbb{R}}}\def\proofend{$\hfill{\Box}$}
\begin{document}

\title{A coherence monotone from Kirkwood-Dirac nonclassicality with respect
to mutually unbiased bases}

\author{Yan Liu}
%\email{guozhihua@snnu.edu.cn}
\affiliation{School of Mathematics and Statistics, Shaanxi Normal University, Xi'an 710119, China}

\author{Zhihua Guo}
\email{guozhihua@snnu.edu.cn}
\affiliation{School of Mathematics and Statistics, Shaanxi Normal University, Xi'an 710119, China}

\author{Zhihao Ma}
\email{mazhihao@sjtu.edu.cn}
\affiliation{School of Mathematical Sciences, MOE-LSC, Shanghai Jiao Tong University, Shanghai 200240, China}
\affiliation{Shanghai Seres Information Technology Co., Ltd, Shanghai 200040, China} \affiliation{Shenzhen Institute for Quantum Science and Engineering, Southern University of Science and Technology, Shenzhen 518055, China}

\author{Shao-Ming Fei}
\email{feishm@cnu.edu.cn}
\address{School of Mathematical Sciences, Capital Normal University, Beijing 100048, China}

\begin{abstract}The Kirkwood-Dirac distribution, serving as an informationally complete representation of a quantum state, has recently garnered increasing attention.
We investigate the Kirkwood-Dirac classicality with respect to mutually unbiased bases.
For prime dimensional Hilbert spaces, {we demonstrate that a quantum state  exhibits Kirkwood-Dirac classicality for two distinct sets of mutually unbiased bases $(A,B)$ and $(A,B')$ if and only if  it is incoherent with respect to $A$}. We subsequently introduce a coherence monotone based on Kirkwood-Dirac nonclassicality with respect to mutually unbiased bases. Additionally, we establish that this coherence monotone can be expressed through weak values, suggesting that quantum coherence can be utilized to detect anomalous weak values.
\end{abstract}

\maketitle

\section{Introduction}
It is widely acknowledged that the noncommutativity inherent in quantum mechanics renders it incompatible with classical physics \cite{00,01}. Quasiprobability distributions are employed to characterize quantum states associated with two noncommutative observables. In contrast to classical probability distributions, quasiprobability distributions can exhibit counterintuitive negativity or even non-reality. This characteristic is believed to confer an advantage in numerous quantum information applications, surpassing the capabilities of any classical theory. For example, the presence of negative Wigner functions manifests quantum advantages in quantum computing and quantum metrology \cite{11,Noah,Liu,Schmid2}.

The Kirkwood-Dirac (KD) distribution \cite{1,2} is yet another distinct quasiprobability distribution that has recently captured considerable attention. {In contrast to the Wigner function, the KD distribution can yield not only negative values but also non-real ones.} These values are termed nonclassical as they cannot be accounted for by classical probability theory. {Instead, they manifest within quantum mechanical probability distributions, such as in the phase space representation of quantum mechanics and quantum optics.} To date, the KD distribution has proven to be pivotal across a broad field of quantum science and technology, spanning from quantum chaos \cite{3,4,5,6,7,8} to quantum metrology \cite{Noah,9,10,11} and the foundations of quantum theory {\cite{13,Zhu,B1+,B2+,B3+,B4+,B5+, Wagner233}}.

The nonclassicality of KD distribution is considered to arise from the paired noncommutativity of operators.  Recently, there has been a proliferation of studies focused on characterizing KD nonclassicality \cite{25,new,26,27,29}. Arvidsson-Shukur \emph{et al.} \cite{25} established the sufficient conditions for KD distribution to be nonclassical and also quantified the KD nonclassicality achievable under different conditions. Recently, Budiyono suggested a sufficient condition for KD nonreality and nonclassicality \cite{new}. De Bi{\`e}vre \cite{26} proved that under the condition that two observables are completely incompatible, only the states with minimum support uncertainty can be KD classical, while the rest are KD nonclassical. Xu \cite{27} established the general structure of KD classical pure states, and explored the connection between the number of the zeros in the transition matrix and the support uncertainty. Langrenez \emph{et al.} \cite{29} characterized how the full convex set of states with positive KD distributions depends on the eigenbases of two observables.  Budiyono \emph{et al.} \cite{B1+} defined a coherence quantifier based on the KD nonclassicality which contains the imaginary and negativity parts of the KD distribution,  called KD-nonclassicality coherence.

As the simplest manifestation of quantum superposition, quantum coherence is a feature of the quantum state associated with a fixed basis, serving as a critical resource in the realm of quantum information and computation \cite{I2,I3,I4,I5,I51,Wagner2222}. In the framework \cite{I6} of resource theory of quantum coherence,  many quantifications of quantum coherence have been extensively explored \cite{I7,I8,I9,I10,I11,I12,I13}.  To establish connections between coherence theory and the KD distribution, Budiyono \emph{et al.} in \cite{32} quantified the coherence of a quantum state using the KD quasiprobability, termed KD coherence.   However, the KD coherence measure may not universally serve as an effective coherence measure, since it is not sure about monotonicity under incoherent operations, which is a crucial criterion for a coherence measure to be deemed appropriate \cite{I14}.  Therefore, it is a worthwhile endeavor to explore the possibility of quantifying coherence in higher dimensional systems through the lens of the KD distribution, firmly grounded within the established framework of the resource theory of coherence.

In this work we delve into the characterization of KD classical states and examine the interrelations between the KD nonclassicality, quantum coherence  and weak values from the measurements of mutually unbiased bases for prime dimensional quantum systems. In Sec. II, we review the definition of KD classical states. We characterize the set of incoherent states in terms of  the KD classicality with respect to (w.r.t.) mutually unbiased bases (MUBs). In Sec. III, we propose a kind of coherence monotone via KD nonclassicality w.r.t. MUBs. As an application, in Sec. IV we present the relationship among the coherence monotone, weak values and the witness of anomalous weak values. We summarize and discuss in Sec. V.

\section{Coherence based on KD nonclassicality w.r.t. mutually unbiased bases}

Consider a finite-dimensional Hilbert space $H$ with dimension $\dim(H)=d$, and let $\mathcal{D}$ denote the set comprising all quantum states on $H$. Let $A=\{|a_m\>\}^{d-1}_{m=0}$ and $B=\{|b_n\>\}^{d-1}_{n=0}$ be two orthonormal bases of $H$. { The families of corresponding projectors of $A$ and $B$ are denoted by $\mathcal{A}=\{\mathcal{A}_m=|a_m\>\<a_m|\}^{d-1}_{m=0}$ and $\mathcal{B}=\{\mathcal{B}_n=|b_n\>\<b_n|\}^{d-1}_{n=0}$, respectively.}

The KD distribution \cite{25} of $\rho\in \mathcal{D}$ w.r.t. $A$ and $B$ is defined by the $d\times d$ matrix $Q^{AB}(\rho)=[Q^{AB}_{mn}(\rho)]$ with the $(m,n)$-entry,
\begin{eqnarray}
Q^{AB}_{mn}(\rho)=\<a_m|\rho|b_n\>\<b_n|a_m\>.
\end{eqnarray}
 The quantum state $\rho$ is said to be {\it KD classical w.r.t. { $(A,B)$}} if $Q^{AB}(\rho)$ is a matrix with nonnegative entries, that is, $Q^{AB}_{mn}(\rho) \geq 0$ for all $m$ and $n$. Otherwise, $\rho$ is said to be {\it KD nonclassical w.r.t.\,{$(A,B)$}}. Denote by $KDC(A,B)$ the set of all KD classical states w.r.t.\,{$(A,B)$}. As per the definition, $KDC(A,B)$ is a compact convex subset of $\mathcal{D}$.

A state is said to be \textit{incoherent} w.r.t. a basis $A$ if it is diagonal under $A$. Denote by $\mathcal{I}(A)$ the set of all incoherent states w.r.t. $A$. It is easy to see that $\mathcal{I}(A)\subset KDC(A,B)$ for any basis $B$. An intriguing inquiry arises: Can all the states that are incoherent w.r.t. basis $A$ be characterized by quantum states that are KD classical with respect to a fixed basis $A$ and a finite set of bases $B$s in $H$? We affirmatively answer this question below for prime-dimensional systems, indicating that the finite set of bases $B$s consists precisely of two mutually unbiased bases with $A$.

In recent years, {MUBs, as widely used measurement bases, play a crucial role in a variety of quantum information tasks such as quantum key distribution, quantum state reconstruction and quantum error correction codes\cite{z1,d1,MUB+1,MUB+2,MUB+3,MUB+4, 33,30}. For a Hilbert space with prime dimension $d$,} there exists a set of $d+1$ bases which are mutually unbiased pairwise. In the case of qubit system ($d = 2$), a set of usual complete MUBs is
{ \begin{eqnarray}
  A&=& \{|a_0\>, |a_1\>\}, \label{10}\\
  B_1&=&\left\{\frac{1}{\sqrt{2}}(|a_0\>+|a_1\>),   \frac{1}{\sqrt{2}}(|0\>-|1\>)\right\}, \label{11}\\
  B_2&=&\left\{\frac{1}{\sqrt{2}}(|a_0\>+{\rm{i}}|a_1\>),   \frac{1}{\sqrt{2}}(|0\>-{i}|1\>)\right\}.  \label{12}
\end{eqnarray}
For $d\geq3$, the $d+1$ MUBs are given by\cite{z1},
\begin{eqnarray}
  A &=& \{|a_0\>, |a_1\>, \ldots, |a_{d-1}\>\}, \label{6}\\
  B_r&=&\{|b^{r}_0\>, |b^{r}_1\>, \ldots, |b^{r}_{d-1}\>\}, r=1,2,\ldots,d  \label{7}
\end{eqnarray}
where $|b^{r}_p\>=(b^{r}_p)_0|a_0\>+(b^{r}_p)_1|a_1\>+\ldots+(b^{r}_p)_{d-1}|a_{d-1}\>,$ $(b^{r}_p)_q=\frac{1}{\sqrt{d}}\omega^{rq^{2}+pq}, ~\omega=\textmd{e}^{\frac{2\pi{i}}{d}},$ $p, q=0, 1, \ldots, d-1$.}

In the following, the notation ``$X\bigcap Y$" denotes the overlap of the two sets $X,Y$. We will discuss the overlap between $KDC(A,B_j)$ and  $KDC(A,B_k)$ for two distinct families of mutually unbiased bases { $(A,B_j)$} and { $(A,B_k)$, respectively}. Our first result is
as follows:

{\bf Theorem 1.} When $d$ is a prime, a state is $A$ incoherent if and only if it is KD classical for two distinct sets of mutually unbiased bases { $(A,B_j)$} and { $(A,B_k)$}, namely,
\begin{equation}
\label{T21}
KDC(A,B_j)\bigcap KDC(A,B_k)=\mathcal{I}(A),
\end{equation}
where $A,B_j,B_k$ $(j\ne k)$ are MUBs defined by either eqs. (\ref{10})-(\ref{12}) or eqs. (\ref{6}), (\ref{7}).

 The proof  of Theorem 1 can be found in Appendix A. Theorem 1 stipulates that quantum states which demonstrate KD classicality w.r.t. both { $(A,B_j)$} and { $(A,B_k)$} must necessarily be incoherent w.r.t. $A$. Conversely, a coherent state w.r.t. $A$ must exhibit KD nonclassicality w.r.t. $A$ and certain $B$ in eqs.\,(\ref{11})-(\ref{12}) or (\ref{7}). Using the result already proven in \cite{29}: if $d = 2$ and $\min_{m,n}{|\langle a_m|b_n\rangle|} > 0$, then  states in $KDC(A,B)$ are just convex combinations of $|a_m\rangle\langle a_m|$ and $|b_n\rangle\langle b_n|(m,n\in \{0,1\})$, we provide a geometric interpretation of Theorem 1 in the two-dimensional case by Figure 1, which illustrates that when $d=2$ and $A=\{|0\>,|1\>\}$, all incoherent states w.r.t. $A$ can be delineated by the quantum states that are KD classical w.r.t fixed bases $(A,B_1)$ and $(A,B_2)$ in eqs. (\ref{10})-(\ref{12}).
\begin{figure}[!htp]
    \centering
    \includegraphics[width=2.5in]{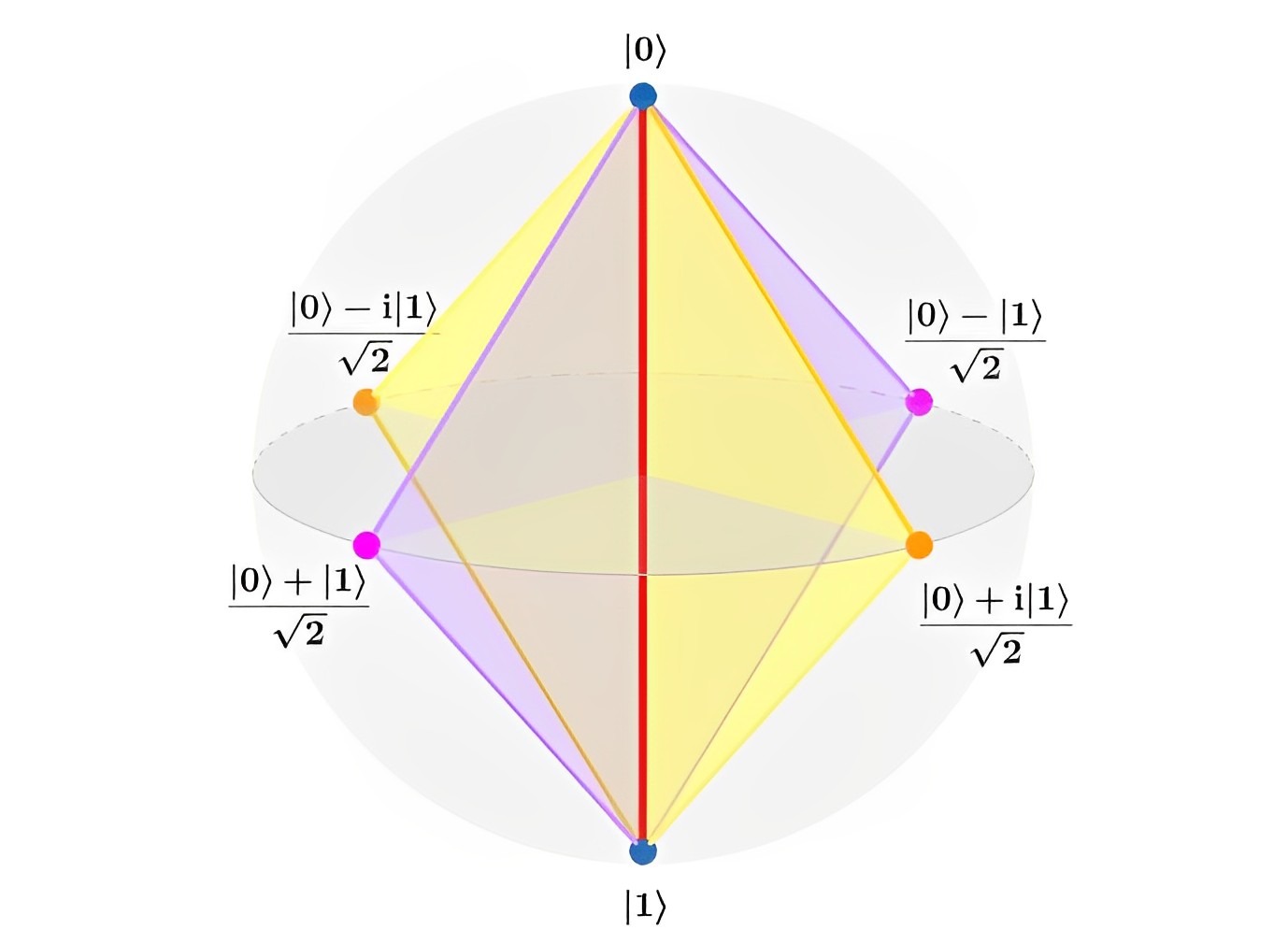}
    \caption{For the case of $d=2$, the purple and yellow quadrilaterals in the Bloch ball depict { $KDC(A,B_1)$ and $KDC(A,B_2)$}, respectively. The red line represents their overlap, which corresponds to the set of all incoherent states w.r.t. $A$.}
    \label{figure1}
\end{figure}

{\it Remark} We remark on our theorem from the perspective of wave-particle dualities. In the experiment of multipath interferometer, the capability of distinguishing which-path the particle took represents the $d$-path distinguishability ${D}_{d}$. The normalized coherence ${C}_d=\sum_{i \neq j} |\rho_{i j}|/(d-1)$ given by the $l_1$-norm coherence measure represents the wave aspect of a particle. They satisfy the following trade-off relation\cite{QureshiD}, ${C}^2_{d} + {D}^2_{d}  \leq 1$, which has been demonstrated in a quantum delayed-choice experiment on large-scale silicon-integrated multipath interferometers\cite{wangjw}. Hence from the perspective of wave-particle duality, if a particle's state is both KD classical w.r.t.\,{$(A,B_j)$ and $(A,B_k)$}, then this particle has vanishing coherence w.r.t. $A$, and thus zero wave property.
Namely, two different MUBs are enough to determine that a particle is classical in the sense that it behaves truly like a `particle'.

\section{A coherence monotone based on KD nonclassicality}

\setcounter{equation}{7}

Any proper coherence measure $C:\mathcal{D}\rightarrow \mathbb{R}$ for a fixed basis should satisfy the following four conditions\cite{I6,I14}:

(e1)  Faithfulness: ${C}(\rho)\geq0$ in general, with equality if and only if $\rho\in \mathcal{I}(A)$.

(e2) Nonincreasing under free operations: $C(\Phi(\rho))\leq C(\rho)$ for any free operation $\Phi$.

(e3) Nonincreasing under free operations on average: $\sum_r p_r C(\rho_r)\leq C(\rho )$ for any free operation $\Phi$ with Kraus operators $\{K_r\}_r$, where $p_r=\textmd{tr}(K_r\rho K_r^\dag)\ne 0$ and $\rho_r=K_r\rho K_r^\dag/p_r$.

(e4) Convexity: ${C}(\sum_{r}p_r\rho_r)\leq\sum_{r}p_r {C}(\rho_r)$ for any quantum state $\rho_r\in \mathcal{D}$ and any probability distribution $\{p_r\}$.

Similar to entanglement monotone \cite{04}, the function $C$ is referred to as a \emph{coherence monotone} if it only satisfies conditions (e1) and (e2). Although a coherence monotone might not adhere to (e3), it can still serve crucial roles in various quantum information processing tasks \cite{I6,I13}.

Recently, Budiyono and Dipojono \cite{32} utilized the KD distribution to access the quantum coherence in a quantum state. Consider an arbitrary quantum state $\rho$ and a basis $A=\{|a_m\>\}^{d-1}_{m=0}$  for a  Hilbert space $H$. The KD coherence of $\rho$ w.r.t. $A$ is defined by
\begin{eqnarray}
C_{KD}[\rho;A]:=\max_{B\in \mathcal{O} }\sum_{m,n}\left|\textmd{Im}\left(Q^{AB}_{mn}(\rho)\right)\right|,\label{5}
\end{eqnarray}
where $\mathcal{O}$ denotes the set of all orthonormal bases for the Hilbert space $H$ of dimensional $d$. It is proven that $C_{KD}[\rho;A]=C_{\ell_1}[\rho;A]$ when $d=2$, where $C_{\ell_1}[\rho;A]$ is the measure of $\ell_1$-norm coherence. Generally the KD coherence is not necessarily a proper coherence measure. Moreover, in (\ref{5}) one needs to estimate the coherence by maximizing over all possible bases $B$. Budiyono also gave coherence quantifiers based on the KD nonclassical values in \cite{58} and admitted a closed expression in terms of nonadditive entropies, in which the optimal values are indeed attained by mutually unbiased bases. Inspired by this result, we will show that, to give a bona fide coherence monotone for prime $d$, {  it is sufficient to take into account much fewer bases and certain kind of incoherent operations (free operations).

Let $A=\{|a_m\>\}^{d-1}_{m=0}\in\mathcal{O}.$ Denote by $\mathcal{F}(A)$ the set of all orthonormal bases for $H$ that are mutually unbiased with $A$, that is,
$$\left\{\{|{b_n}\>\}_{n=0}^{d-1}\in \mathcal{O}:|{b_n}\>=\frac{1}{\sqrt{d}}
\sum_{k=0}^{d-1}\textmd{e}^{\textmd{i}\theta_{nk}}|a_k\>, \theta_{nk}\in [0,2\pi)\right\}. $$
 Clearly, all $B_r$s defined by (\ref{7}) belong to $\mathcal{F}(A)$.

In \cite{35}, E. Chitambar and G. Gour introduced the class of physically incoherent operations (PIOs), which represents
the coherence analog to LOCC for a physically
consistent resource theory. They proved that PIOs are special strictly incoherent operations and  every PIO  can be expressed as a convex combination of operations with Kraus operators $\{K_r\}_{r=1}^q$ of the form $K_r=U_rP_r$, where $U_r=\sum_{x\in S_r}\textmd{e}^{{i}\theta_{x}}|a_{\pi_r(x)}\>\<a_x|$, $P_r=\sum_{x\in S_r}|a_x\>\<a_x|$, $\pi_r$ is a permutation of $S_r$, i.e., a bijection from $S_r$ onto itself, $\{S_1,S_2,\ldots,S_q\}$ is a partition of the index set $\{0,1,\ldots,{d-1}\}$. It is easy to see that
\begin{equation}\label{UP}
U_rP_r=\sum_{x\in S_r}\textmd{e}^{{i}\theta_{x}}|a_{\pi_r(x)}\>\<a_x|(r=1,2\ldots,q).
\end{equation}
We call PIOs with the Kraus operators of the form (\ref{UP}) the UP-PIOs.

{\bf Theorem 2.} When $d$ is prime, the following KD nonclassicality based quantity,
\begin{eqnarray}\label{thm2}
\widehat{C}_{KD}[\rho;A]:=\max_{B\in \mathcal{F}(A)}\sum_{m,n}\left|\textmd{Im}\left(Q^{AB}_{mn}(\rho)\right)\right|
\end{eqnarray}
is a convex coherence monotone.}

We show that $\widehat{C}_{KD}$ satisfies (e1), (e4) and (e2) under PIOs, i.e.,
$\widehat{C}_{KD}[\Phi_{PIO}(\rho);A]\leq\widehat{C}_{KD}[\rho;A]$. See the proof in Appendix B. Moreover, for qubit case, $\widehat{C}_{KD}$ reduces to the $\ell_1$-norm coherence measure. In addition, the same proof applies for the monotonicity of KD coherence $C_{KD}$ proposed by Budiyono and Dipojono in \cite{32} under PIOs.

Let us exemplify the coherence monotone $\widehat{C}_{KD}$ by two examples for qutrit case. Set $A=\{|0\>,|1\>,|2\>\}$. Consider the state $\rho=|\psi\>\<\psi|$ with $|\psi\>=\lambda_0|0\>+\lambda_1|1\>+\lambda_2|2\>$, where $\lambda_0^2+\lambda_1^2+\lambda_2^2=1$.
Observing Figure 2, it is evident that $\widehat{C}_{KD}[\rho;A]\leq {C}_{\ell_1}[\rho;A]$, and the inequality is saturated when one of the parameters $\lambda_0$, $\lambda_1$ or $\lambda_2$ equals 0. The maximum $\frac{2}{\sqrt{3}}$ of $\widehat{C}_{KD}[\rho;A]$ is attained when $|\lambda_0|=|\lambda_1|=|\lambda_2|=\frac{1}{\sqrt{3}}$.
\begin{figure}[!htp]
    \centering
    \includegraphics[width=2.5 in]{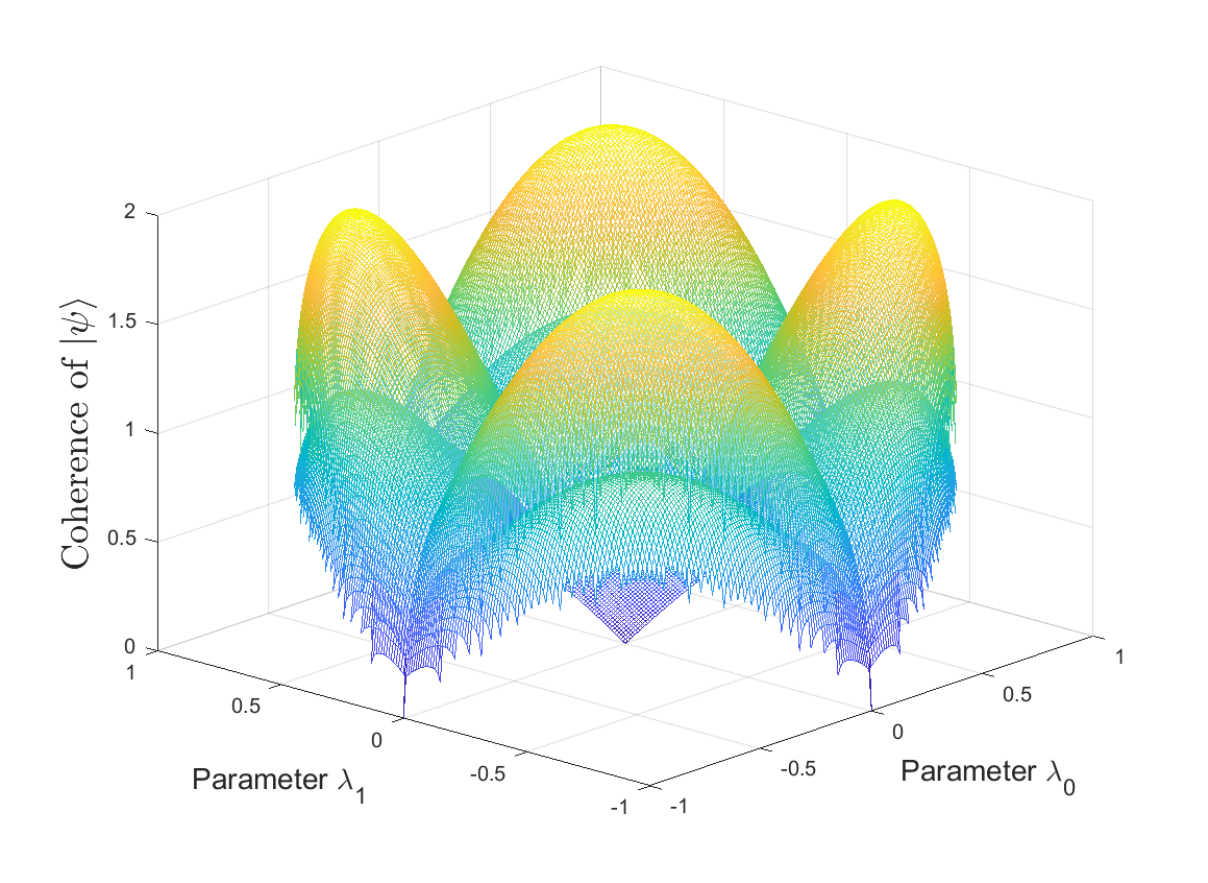}
    \caption{ The bottom (top) surface represents $\widehat{C}_{KD}[\rho;A]$ ($C_{\ell_1}[\rho;A]$) versus parameters $\lambda_0$ and $\lambda_1$.}
    \label{figure1}
\end{figure}

{For the qutrit mixed state $\sigma=\frac{1}{2}(|0\>\<0|+|1\>\<1|)+\mu(|0\>\<1|+|1\>\<0|)$ with $0\le \mu\le \frac{1}{2}$,
$\widehat{C}_{KD}[\sigma;A]= \frac{1}{2}\mu\max_{[\theta_{st}]\in \Gamma}\left\{|\sin(\theta_{00}-\theta_{01})|
+|\sin(\theta_{11}-\theta_{10})|\right\}
=\frac{\sqrt{3}}{2}\mu.
$

 The coherence monotone $\widehat{C}_{KD}[\rho;A]$ presented in Theorem 2 tightly associates with quantum uncertainty relations. The Heisenberg uncertainty relation says for any two observables $X$ and $Y$, one has
$$
\Delta(X)\Delta(Y)\geq \frac{1}{2}|\tr([X,Y]\rho)|=\left|\textmd{Im}(\tr(\rho XY))\right|
$$
associated with a state $\rho$, where $\Delta(X)=\sqrt{\textmd{tr}(X^2\rho)-(\textmd{tr}X\rho)^2}$. When the eigenvectors $\{|a_m\rangle\}$ and $\{|b_n\rangle\}$ of $X$ and $Y$ constitute a pair of mutually unbiased bases, the incompatibility $\max_{m,n}|\<a_m|b_n\>|$ of $X$ and $Y$ attains the maximum $\frac{1}{\sqrt{d}}$. In this context, the lower bound for the aggregate sum of variances' products across all spectrum projections is precisely characterized by the $\ell_1$-norm of the imaginary component of KD distribution of $\rho$ w.r.t. $\{|a_m\rangle\}$ and $\{|b_n\rangle\}$.
Consider the fixed basis $A=\{|a_m\rangle\}$. We naturally seek to maximize over any basis $\{|b_n\rangle\}$ in all possible mutually unbiased bases with $A$, aiming to optimize $\sum_{m}\Delta(|a_m\rangle\langle a_m|)\sum_{n}\Delta(|b_n\rangle\langle b_n|)$. The lower bound of this maximization is precisely defined by the coherence monotone $\widehat{C}_{KD}[\rho;A]$, that is,
$$\widehat{C}_{KD}[\rho;A]\le\sum_{m}\Delta(|a_m\rangle\langle a_m|)\max_{B\in\mathcal{F}(A)}\sum_{n}\Delta(|b_n\rangle\langle b_n|).$$

\section{Connections between coherence and weak values}

As an application, let us explore the interconnection between the coherence monotone $\widehat{C}_{KD}$ and quantum mechanical weak values. In 1988, Aharonov, Albert and Vaidman proposed a weak measurement scheme \cite{36}, which involves preparing a state $|\psi\>$, followed by a weak interaction between the system and a measurement apparatus, generated by some observable $O$, with a final postselection onto some state $|\phi\>$. The average change in the apparatus pointer is given by the weak value: $O_{\textit{w}}=\frac{\<\phi|O|\psi\>}{\<\phi|\psi\>}.$ The weak value $O_{\textit{w}}$ is called {\textit{anomalous}} if it lies outside the range of the spectrum $\sigma(O)$ of $O$ \cite{361}. The weak value has gained entrance to a more public spotlight due to its experimental applications \cite{362,363,364,Wagner220}. Recently, Wagner and Galv\~{a}o expanded the definition of weak value to encompass scenarios where preselected state $\rho$ and postselection state $\sigma$ involve mixed states as
$O_{\textit{w}}=\frac{\textmd{tr}(\sigma O \rho )}{\textmd{tr}(\sigma \rho )},$ and verified that the anomalous weak values imply that the pre- and postselection states have coherence in the eigenbasis of $O$, but  in
general coherence of $\rho$ and $\sigma$ does not imply anomaly of $O_{\textit{w}}$ \cite{37}.

Let $A=\{|a_m\>\}$ be an orthonormal basis of $H$ with prime dimension. Interestingly, the coherence monotone $\widehat{C}_{KD}[\rho;A]$ can be represented by certain weak values,
\begin{eqnarray*}
\widehat{C}_{KD}[\rho;A]&=&\max_{B\in \mathcal{F}(A)}\sum_{m,n}\left|\textmd{Im}\left(Q^{AB}_{mn}(\rho)\right)\right|\\
&=&\max_{\{|b_n\>\}\in\mathcal{F}(A)}\sum_{m,n}
\left|\textmd{Im}\frac{\textmd{tr}(\mathcal{B}_n\mathcal{A}_m\rho)}
{\textmd{tr}(\mathcal{B}_n\rho)}\right|\textmd{tr}(\mathcal{B}_n\rho)\\
&=&\max_{\{|b_n\>\}\in\mathcal{F}(A)}
\sum_{m,n}|\textmd{Im}(\mathcal{A}_{m}^{\textit{w}}(\rho,\mathcal{B}_n))|\textmd{tr}(\mathcal{B}_n\rho),
\end{eqnarray*}
where $\mathcal{A}_{m}^{\textit{w}}(\rho,\mathcal{B}_n)$ is the weak value of $\mathcal{A}_m=|a_m\>\<a_m|$ with the preselected state $\rho$ and postselection state $\mathcal{B}_n=|b_{n}\>\<b_n|$. Furthermore, we present the following theorem, The proof of which is detailed in Appendix C.

{\bf Theorem 3.}  For any prime dimensional quantum system and a fixed basis $A=\{|a_m\>\}$, a state $\rho$ is coherent if and only if there exists some $\mathcal{A}_m=|a_m\>\<a_m|$ and  $\mathcal{B}_n=|b_{n}\>\<b_n|$ such that the weak value $\mathcal{A}_{m}^{\textit{w}}(\rho,\mathcal{B}_n)$  is nonreal and therefore anomalous, where $\{|b_n\>\}\in\mathcal{F}(A)$.

Theorem 3 above establishes that the existence of an anomalous $\mathcal{A}_{m}^{\textit{w}}(\rho,\mathcal{B}_n)$ is equivalent to the coherence of $\rho$. This result also shows that the coherence monotone $\widehat{C}_{KD}$ is a witness of anomalous weak values.

\section{Conclusion}

We have explored the characteristics of coherence based on  KD nonclassicality with respect to two bases $(A,B)$. Specifically, when $d$ is a prime, our analysis reveals that the ensemble of incoherent states precisely corresponds to the overlap of two sets of KD classical states with respect to two distinct families of mutually unbiased bases { $(A,B)$ and $(A,B')$}, respectively.

Furthermore, we have derived a coherence monotone induced by KD nonclassicality
{ $$\widehat{C}_{KD}[\rho;A]=\max_{B\in\mathcal{F}(A)} \sum_{m,n}\left|\textmd{Im}\left(Q^{AB}_{mn}(\rho)\right)\right|,$$
where $\mathcal{F}(A)$ contains all MUBs corresponding to $A$.} It has been demonstrated  that $\widehat{C}_{KD}$ fulfills all the requisite properties of a coherence monotone. Specifically for qubit states, $\widehat{C}_{KD}$ equates to the $\ell_1$ norm coherence measure. Additionally, we have illustrated the computational tractability of coherence monotone $\widehat{C}_{KD}$ through two examples in the qutrit scenario.  In addition, we have highlighted certain connections between the coherence monotone $\widehat{C}_{KD}[\rho;A]$ and quantum uncertainty relations.
Finally, we have studied the relationship between the coherence monotone and the quantum mechanical weak values. Our investigation manifests that our coherence monotone $\widehat{C}_{KD}$ can be elucidated through weak values, certifying that quantum coherence can witness anomalous weak values.

\vskip 0.3cm
\noindent {\bf Acknowledgments}.  This work was supported by  the National Natural Science Foundation of China (Grant Nos.\,12271325, 12371132, 12075159 and 12171044) and the Academician Innovation Platform of Hainan Province.
%the Beijing Postdoctoral Research Foundation,

\vskip 0.3cm

%\end{document}

\section*{Appendix A. Proof of Theorem 1}
\renewcommand{\theequation}{A.\arabic{equation}}
\setcounter{equation}{0}

According to the Theorem 1.1 in Ref. \cite{29}, one sees that the convex combinations of the eigenprojectors of $A$ and $B$ constitute the only KD classical states, i.e., $$\textmd{conv}\{|a_m\rangle\langle a_m|,|b_n\rangle\langle b_n|:m,n=0,1,\ldots,d-1\}$$ under any one of the following hypotheses:

(i) $d = 2$ (for qubits) and $m_{A,B}=\min_{m,n}{|\langle a_m|b_n\rangle|} > 0$;

(ii) $d$ is prime and the transition matrix $U^{AB}$ from $A$ to $B$ is the discrete Fourier transform (DFT) matrix.

When $d=2$, if $m_{A,B}=\min_{m,n}{|\langle a_m|b_n\rangle|} > 0$, then it follows from (i) that $KDC(A, B_1)$ = conv({$\mathcal{A}\bigcup\mathcal{B}_1$}),
$KDC(A, B_2$) = conv({$\mathcal{A}\bigcup\mathcal{B}_2$}). It then follows that conv({$\mathcal{A}\bigcup\mathcal{B}_1$}) and conv({$\mathcal{A}\bigcup\mathcal{B}_2$})
are rectangles in the Bloch sphere with a common diagonal. Hence $KDC(A, B_1)\bigcap KDC(A, B_2) =\mathcal{I}(A)$.

When $d$ is an odd prime number, the $d+1$ mutually unbiased bases are defined as the form of eqs.\,(5),\,(6).\,We introduce the auxiliary basis $A'=\{|a_j'\rangle:|a_j'\rangle=\omega^{rj^2}|a_j\rangle\}$. It is clear that $KDC(A, B_r) = KDC(A', B_r)(r=1,2,\ldots,d)$.
Since $$\langle a_j' |b_p^r\rangle=\frac{1}{\sqrt{d}}\omega^{pj}, p,j=0,1,\ldots,d-1,$$
the transition matrix between $A'$ and $B_r$ is the DFT matrix. Hence, from (ii), one finds
$$KDC(A, B_r) =\textmd{conv}{ \{\mathcal{A}'\bigcup \mathcal{B}_r\}=\textmd{conv}\{\mathcal{A}\bigcup\mathcal{B}_r\}}.$$

Suppose that $\rho$ is KD classical for two distinct sets of mutually unbiased bases {$(A,B_j)$ and $(A,B_k)$}, i.e., $\rho\in KDC(A,B_j)\bigcap KDC(A,B_k)$. Then $\rho$ must be a convex mixture of the basis states of $\mathcal{A}$ and $\mathcal{B}_j$, and also be a convex mixture of the basis states of $\mathcal{A}$ and $\mathcal{B}_k$, i.e.,
\begin{eqnarray*}
\rho&=&\sum^{d-1}_{s=0}\lambda_s|a_s\>\<a_s|+\sum^{d-1}_{t=0}\mu_t|b^j_t\>\<b^j_t|
\\
&=&\sum^{d-1}_{s=0}\alpha_s|a_s\>\<a_s|+\sum^{d-1}_{t=0}\beta_t|b^k_t\>\<b^k_t|,
\end{eqnarray*}
where $\sum^{d-1}_{s=0}\lambda_s+\sum^{d-1}_{t=0}\mu_t=1$ and $\sum^{d-1}_{s=0}\alpha_s+\sum^{d-1}_{t=0}\beta_t=1$.
Combining with the fact that $\tr(\rho)=1$, we compute that
\begin{eqnarray*}
\<b^j_n|\rho|b^j_n\>&=&\sum^{d-1}_{s=0}\lambda_{s}|\<a_s|b^j_n\>|^2+\mu_{n}\\
&=&\sum^{d-1}_{s=0}\alpha_{s}|\<a_s|b^j_n\>|^2+\sum^{d-1}_{t=0}\beta_{t}|\<b^k_t|b^j_n\>|^2
\end{eqnarray*}
for each $n=0,1,2,\ldots,d-1$.
Since $A$, $B_1$ and $B_2$ are mutually unbiased, $|\<a_s|b^j_n\>|^2=|\<b^k_t|b^j_n\>|^2=\frac{1}{d}$ for any $s,n,t$. Then for any $n$,
$$\frac{\sum^{d-1}_{s=0}\lambda_{s}}{d}+\mu_{n}=\frac{\sum^{d-1}_{s=0}\alpha_{s}
+\sum^{d-1}_{t=0}\beta_{t}}{d}=\frac{1}{d},$$
which implies $\sum^{d-1}_{s=0}\lambda_{s}+d\mu_{n}=1$ for any $n$. Based on $\sum_{s=0}^{d-1}\lambda_{s}+\sum_{t=0}^{d-1}\mu_{t}=1$, we have $d\mu_{n}=\sum_{t=0}^{d-1}\mu_{t}$  for any $n$. This implies that  $\mu_0=\mu_1=\ldots=\mu_{d-1}$ and  so
\begin{eqnarray*}
\rho=\sum^{d-1}_{s=0}(\lambda_s+\mu_0)|a_s\>\<a_s|\in\mathcal{I}(A).
\end{eqnarray*}
Thus, $KDC(A,B_j)\bigcap KDC(A,B_k)\subseteq\mathcal{I}(A)$ from the arbitrariness of $\rho$.
Conversely, for any $B_j$ defined in (6) and $\rho=\sum^{d-1}_{s=0}\rho_{ss}|a_s\>\<a_s|\in \mathcal{I}(A)$, $Q_{mn}^{AB_j}(\rho)=\frac{\rho_{mm}}{d}\geq 0(\forall m,n)$  can be obtained by direct calculation, which implies that $\rho$ is KD classical w.r.t.\,$(A,B_j)$. This implies that $\mathcal{I}(A)\subseteq KDC(A,B_j)\bigcap KDC(A,B_k).$ Hence, $KDC(A,B_j)\bigcap KDC(A,B_k)=\mathcal{I}(A).$

In summary, when $d$ is a prime, $A,B_j,B_k$ $(j\ne k)$ are defined by either eqs.\,(2)-(4) or eqs.\,(5),\,(6), quantum states which are both KD classical w.r.t. {$(A,B_j)$ and $(A,B_k)$} must be  incoherent w.r.t.$A$. $\hfill\square$

\vskip 0.6cm
\noindent {\bf Appendix B. Proof of Theorem 2}

To prove the property (e1), first note that if $\rho \in \mathcal{I}(A)$, then $Q^{AB}_{mn}(\rho)\geq0$ for all $B\in\mathcal{F(A)},m,n\in[d]$. Hence, $\widehat{C}_{KD}[\rho;A]=0$ by definition.

Conversely, if $\widehat{C}_{KD}[\rho;A]=0$, then $Q^{AB}_{mn}(\rho)\in \R$  for all $B\in\mathcal{F(A)},m,n\in[d]$. Especially,  $Q^{AB_r}_{mn}(\rho)\in \R$ for the basis $B_r$ defined by eq.\,(\ref{7}) and all $m,n\in[d]$, i.e., the KD distribution of $\rho$ w.r.t. { $(A,B_r)$}  is real. Using this conclusion for $B_r=\{|b^r_t\>\}_{t=0}^{d-1}(r=1,2)$ and Proposition 3.2 in Ref. \cite{29} for prime $d$, there must exist nonnegative numbers $\lambda_{1,s}, \mu_{1,t},\lambda_{2,s}, \mu_{2,t}$ such that
$
\rho=\sum_{s=0}^{d-1}\lambda_{1,s}|a_s\>\<a_s|+\sum_{t=0}^{d-1}\mu_{1,t}|b^1_t\>\<b^1_t|=\sum_{s=0}^{d-1}\lambda_{2,s}|a_s\>\<a_s|+\sum_{t=0}^{d-1}\mu_{2,t}|b^2_t\>\<b^2_t|
,$
where $\sum_{s=0}^{d-1}\lambda_{1,s}+\sum_{t=0}^{d-1}\mu_{1,t}
=\sum_{s=0}^{d-1}\lambda_{2,s}+\sum_{t=0}^{d-1}\mu_{2,t}=1$. By using a proof method similar to Theorem 1, one concludes that  $\rho$ is incoherent w.r.t. $A$.

The property (e4) can be proved by directly using the triangle inequality:
{\begin{eqnarray*}
\widehat{C}_{KD}\left[\sum_{k}p_k\rho_k;A\right]&\leq& \sum_{k}p_k\widehat{C}_{KD}[\rho_k;A].
\end{eqnarray*}}

To prove property (e2), we let $\Phi$ be any UP-PIO of $H$, i.e., $\Phi(\rho)=\sum_rK_r\rho K^{\dag}_r$ for all states $\rho$ of $H$, where $K_r=U_rP_r$, $U_r=\sum_{x\in S_r}\textmd{e}^{{i}\theta_{x}}|\pi_r(a_x)\>\<a_x|$,  $P_r=\sum_{x\in S_r}|a_x\>\<a_x|$, $\pi_r$ is a permutation of $S_r$ and $\{S_1,S_2,\ldots,S_q\}$ is a partition of the index set $\{0,1,\ldots,{d-1}\}$.

For any $B\in \mathcal{F}(A)$ and any $m,n$ in $\{0,1,\ldots,{d-1}\}$, let $m\in S_{r_m}$. Then when $r\ne r_m$,  $\<a_{x}|a_{\pi_{r}^{-1}(m)}\>=0$ for all $x\in S_r$; when $r= r_m$,  $\<a_{x}|a_{\pi_{r}^{-1}(m)}\>=0$ for all $x\in S_{r_m}$ with $\pi_{r_m}(x)\ne m$. Hence, we obtain from eq. (\ref{UP}) that
\begin{eqnarray*}
&&\textmd{tr}\left(|b_n\>\<b_n|a_m\>\<a_m|\Phi(\rho)\right) \\
&=&\textmd{tr}\left(|\widetilde{b_n}\>\textmd{e}^{{i}\theta_{\pi_{r_m}^{-1}(m)}}\<b_n|a_m\>\<a_{\pi_{r_m}^{-1}(m)}|\rho\right),\\
\end{eqnarray*}
where
\begin{equation}\label{OO}
|\widetilde{b_n}\>
=
\sum_{y=0}^{d-1}\textmd{e}^{-\textmd{i}\theta_y}\<a_{\pi(y)}|b_n\>|a_y\>,\end{equation}
in which $\pi$ is the bijection on the index set $\{0,1,\ldots,d-1\}$ defined by   $\pi(x)=\pi_r(x)$ if $x\in S_r$.
Note that when $r=r_m$, it holds that
$$
\sum_{x\in S_{r_m}}\textmd{e}^{\textmd{i}\theta_{x}}\<b_n|a_{\pi_{r_m}(x)}\>\<a_{x}|a_{\pi_{r_m}^{-1}(m)}\>=
\textmd{e}^{{i}\theta_{\pi_{r_m}^{-1}(m)}}\<b_n|a_m\>,$$
and when $r\in[q]\setminus\{r_m\}$, it holds that
$\<a_{x}|a_{\pi_{r}^{-1}(m)}\>=0$ for all $x\in S_r$, and so
$$\sum_{x\in S_{r}}\textmd{e}^{\textmd{i}\theta_{x}}
\<b_n|a_{\pi_{r_m}(x)}\>\<a_{x}|a_{\pi_{r_m}^{-1}(m)}\>=0.$$
Thus, we have
$
\textmd{tr}\left(|b_n\>\<b_n|a_m\>\<a_m|\Phi(\rho)\right)
=\textmd{tr}\left(|\widetilde{b_n}\>\<\widetilde{b_n}|a_{\pi^{-1}(m)}\>\<a_{\pi^{-1}(m)}|\rho\right).
$
Next, let us check $\widetilde{B}=\{|\widetilde{b_n}\>\}\in\mathcal{F}(A).$ First, we compute from the definition (\ref{OO}) of $\widetilde{B}$ that
\begin{eqnarray*}
&&\<\widetilde{b_s}|\widetilde{b_t}\>\\
&=&\sum_{r,j=1}^q\sum_{x\in S_j y\in S_r}\textmd{e}^{\textmd{i}\theta_y-\textmd{i}\theta_x}
\<b_s |a_{\pi_r(y)}\>\<a_y| \<a_{\pi_j(x)}|b_t\>|a_x\>\\
&=&\sum_{r=1}^q\sum_{y\in S_r}\<a_{\pi_r(y)}|b_t\>\<b_s |a_{\pi_r(y)}\>\\
&=&\delta_{s,t}.
\end{eqnarray*}
This shows that $\widetilde{B}$ is an orthonormal basis for $H$.

Moreover, let $n\in S_{r_n}$. Then for each $s=0,1,\ldots,d-1$, it holds that $\<\widetilde{b_s}|a_n\>$ equals to
$$\sum_{r=1}^q\sum_{y\in S_r}\textmd{e}^{\textmd{i}\theta_y}\<b_s |a_{\pi_r(y)}\>\<a_y|a_n\>=
\textmd{e}^{\textmd{i}\theta_n}\<b_s |a_{\pi_{r_n}(n)}\>,$$
implying that $|\<\widetilde{b_s}|a_n\>|=|\<b_s |a_{\pi_{r_n}(n)}\>|=\sqrt{\frac{1}{d}}.$
Hence, $\widetilde{B}$ and $A$ are mutually unbiased bases for $H$, i.e.,  $\{|\widetilde{b_n}\>\}_{n=0}^{d-1}\in \mathcal{F}(A)$. So we obtain that
\begin{eqnarray*}
&&\sum_{m,n}\left|\textmd{Im}\left(Q_{mn}^{AB}(\Phi(\rho))\right)\right|\\
&=&\sum_{m,n}\left|\textmd{Im}\left(\textmd{tr}\left(
|\widetilde{b_n}\>\<\widetilde{b_n}|\pi^{-1}(a_m)\>\<\pi^{-1}(a_m)|\rho\right)\right)\right|\\
&=&\sum_{m',n}\left|\textmd{Im}\left(\textmd{tr}\left(
|\widetilde{b_n}\>\<\widetilde{b_n}|a_{m'}\>\<a_{m'}|\rho\right)\right)\right|\\
&\leq&\max_{B\in \mathcal{F}(A) }\sum_{m,n}\left|\textmd{Im}\left(Q_{mn}^{AB}(\rho)\right)\right|\\
  &=&\widehat{C}_{KD}[\rho_;A].
\end{eqnarray*}
This shows that $\widehat{C}_{KD}[\Phi(\rho);A]\le\widehat{C}_{KD}[\rho_;A]$
is valid for every UP-PIO $\Phi$ of $H$.
Since any PIO $\Phi$ of $H$ is a convex combination of UP-PIOs, we conclude from   the property (e4) that  $\widehat{C}_{KD}[\Phi(\rho);A]\leq \widehat{C}_{KD}[\rho_;A].$ \proofend

\vskip 0.6cm
\noindent {\bf Appendix C.  Proof of Theorem 3}
\vskip 0.3cm

{  Suppose that the dimension of the quantum system is prime. Then
\begin{eqnarray*}
\widehat{C}_{KD}[\rho;A]=\max_{
\{|b_n\>\}\in\mathcal{F}(A)}\sum_{m,n}|\textmd{Im}
(\mathcal{A}_{m}^{\textit{w}}(\rho,\mathcal{B}_n))|\<b_n|\rho|b_n\>,
\end{eqnarray*}
where $\mathcal{A}_{m}^{\textit{w}}(\rho,\mathcal{B}_n)=\frac{\textmd{tr}(\mathcal{B}_n\mathcal{A}_m\rho)}
{\textmd{tr}(\mathcal{B}_n\rho)}$ is the weak value of $\mathcal{A}_m=|a_m\>\<a_m|$ with the preselected state $\rho$ and postselection state $\mathcal{B}_n=|b_{n}\>\<b_n|$.

If $\rho$ is coherent w.r.t.\,$A$, then $\widehat{C}_{KD}[\rho;A]>0$, and thus there  exists a basis $\{|b_n\>\}\in\mathcal{F}(A)$ such that $\sum_{m,n}|\textmd{Im}(\mathcal{A}_{m}^{\textit{w}}(\rho,\mathcal{B}_n))|\<b_n|\rho|b_n\>>0,$
which implies that for some $\mathcal{A}_{m}$ and  $\mathcal{B}_{n}$,
$\textmd{Im}(\mathcal{A}_{m}^{\textit{w}}(\rho,\mathcal{B}_n))\ne 0.$
 Hence, the weak value $\mathcal{A}_{m}^{\textit{w}}(\rho,\mathcal{B}_n)$ is anomalous.

Conversely, suppose that there exists the postselected state $\mathcal{B}_n$ such that the weak value $\mathcal{A}_{m}^{\textit{w}}(\rho,\mathcal{B}_n)$ of some $\mathcal{A}_{m}$ is anomalous, where $\{|b_n\>\}\in\mathcal{F}(A)$.  Assume $\rho$ is incoherent w.r.t. $A$, that is, $\rho=\sum_{s}\rho_{ss}|a_s\>\<a_s|$. Then
\begin{eqnarray*}
  \mathcal{A}_{m}^{\textit{w}}(\rho,\mathcal{B}_n)&=& \frac{\<b_n|\mathcal{A}_m(\sum_{s}\rho_{ss}|a_s\>\<a_s|)|b_n\>}{\<b_n|
  (\sum_{s}\rho_{ss}|a_s\>\<a_s|)|b_n\>} = \rho_{mm}.
\end{eqnarray*}
It is obvious that $\rho_{mm}$ is in the range of $\sigma (\mathcal{A}_m)$ since $\sigma (\mathcal{A}_m)=\{0,1\}$. This contradicts with the assumption that the weak value $\mathcal{A}_{m}^{\textit{w}}(\rho,\mathcal{B}_n)$ is anomalous. Thus, $\rho$ is coherent w.r.t. $A$. $\hfill\square$}

\end{document}